\documentclass[a4paper,12pt]{article}

\usepackage[T1]{fontenc} 
\usepackage[utf8]{inputenc} 
\usepackage{fullpage} 
\usepackage[parfill]{parskip} 
\usepackage{setspace}
\usepackage{natbib} 
\usepackage{microtype} 
\usepackage{amsmath} 
\usepackage{tikz}\usepackage{pgfplots}\pgfplotsset{compat=1.16} 
\usepackage{amssymb} 
\usepackage[font=small,labelfont=bf]{caption} 

\title{Lexicographic Enumeration of Set Partitions}
\author{Giorgos Stamatelatos \hspace{1cm} Pavlos S. Efraimidis}
\date{}

\begin{document}

\maketitle

\begin{abstract}\noindent
In this report, we summarize the set partition enumeration problems and thoroughly explain the algorithms used to solve them. These algorithms iterate through the partitions in lexicographic order and are easy to understand and implement in modern high-level programming languages, without recursive structures and jump logic. We show that they require linear space in respect to the set cardinality and advance the enumeration in constant amortized time. The methods discussed in this document are not novel. Our goal is to demonstrate the process of enumerating set partitions and highlight the ideas behind it. This work is an aid for learners approaching this enumeration problem and programmers undertaking the task of implementing it.
\end{abstract}

\section{Introduction}
\label{sec:introduction}

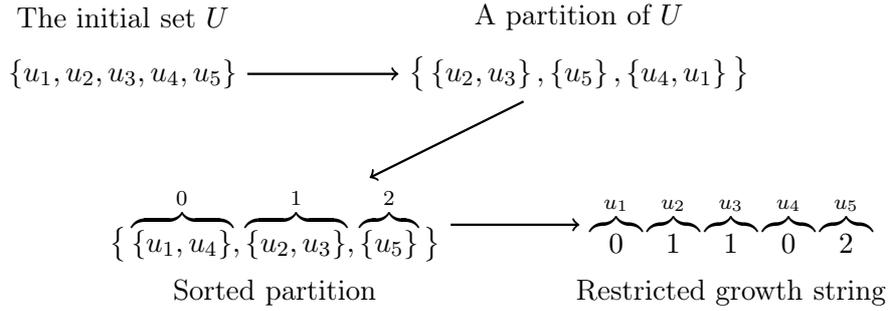
\begin{figure}
    \centering
    \small
    \begin{tikzpicture}
    \node (A) at (-3,1) {$\left\{ u_1,u_2,u_3,u_4,u_5 \right\}$};
    \node (Adesc) at (-3,1.75) {The initial set $U$};
    \node (B) at (3,1) {$\big\{ \left\{u_2,u_3\right\},\left\{u_5\right\},\left\{u_4,u_1\right\} \big\}$};
    \node (Bdesc) at (3,1.75) {A partition of $U$};
    \node (C) at (-1,-1) {$\big\{ \overbrace{\left\{u_1,u_4\right\}}^0,\overbrace{\left\{u_2,u_3\right\}}^1,\overbrace{\left\{u_5\right\}}^2 \big\}$};
    \node (Cdesc) at (-1,-1.9) {Sorted partition};
    \node (D) at (5,-1) {$\overbrace{0}^{u_1} \overbrace{1}^{u_2} \overbrace{1}^{u_3} \overbrace{0}^{u_4} \overbrace{2}^{u_5}$};
    \node (Ddesc) at (5,-1.9) {Restricted growth string};
    \path [->](A) edge[line width=0.3mm] node[left] {} (B);
    \path [->](B) edge[line width=0.3mm] node[left] {} (C);
    \path [->](C) edge[line width=0.3mm] node[left] {} (D);
    \end{tikzpicture}
    \caption{The representation of a partition as a growth string. The blocks of the partition are sorted based on the least element in each block, in this example $u_1$, $u_2$ and $u_5$. Each number $a_i$ in the growth string shows the block containing $u_i$.}
    \label{fig:growth-string}
\end{figure}

\citet[Section 7.2.1.5]{knuth2014art} discusses the problem of partition enumeration, which consists in enumerating the number of ways that a set can be partitioned into non-empty subsets (or blocks). Following Knuth's notation, each partition of a set $U=\{u_1,u_2,\dots,u_n\}$ can be represented by a \textit{restricted growth string} $a_1 a_2 \dots a_n$ such that
\begin{equation}\label{eq:restricted-growth-string}
    a_1 = 0 \text{~~~and~~~} a_{j+1} \le 1 + \text{max}(a_1,\dots,a_j) \text{~for~} 1 \le j < n,
\end{equation}
where $a_z$ shows the block at which $u_z$ is in. Hence, two elements $u_i$ and $u_j$ are in the same block if and only if $a_i = a_j$. For example, the 15 possible partitions of the set with 4 elements are
\begin{quote}
    0000, 0001, 0010, 0011, 0012, 0100, 0101, 0102, 0110, 0111, 0112, 0120, 0121, 0122, 0123.
\end{quote}
These partitions are also presented here in \textit{lexicographic order} in respect to their restricted growth string representation. The partitions in this enumeration can be grouped based on the number of blocks that they have, where a $k$-partition is a partition with exactly $k$ blocks. For example, the above set has a single 1-partition which is the 0000 and a single 4-partition which is the 0123. The 2-partitions, in lexicographic order, are
\begin{quote}
    0001, 0010, 0011, 0100, 0101, 0110, 0111
\end{quote}
and the 3-partitions are, in the same order,
\begin{quote}
    0012, 0102, 0112, 0120, 0121, 0122.
\end{quote}
The analogy of a partition with its restricted growth string is illustrated in Figure~\ref{fig:growth-string}. Although it is possible to set $a_1$ as any arbitrary value higher than 0, such change is inconsequential and, for simplicity, we study the $a_1=0$ case. The restrictions in Equation~\ref{eq:restricted-growth-string} are necessary in order to ensure that each partition has a unique restricted growth string representation; for example 0012 and 0021 that point to the same partition would otherwise both be traversed as part of the enumeration.

\begin{figure}
    \centering
    \small
    \begin{tikzpicture}
    \node[shape=circle,draw=black,line width=0.25mm] (A) at (0,0) {A};
    \node[shape=circle,draw=black,line width=0.25mm] (B) at (0.6,1) {B};
    \node[shape=circle,draw=black,line width=0.25mm] (C) at (1.8,1) {C};
    \node[shape=circle,draw=black,line width=0.25mm] (D) at (1.2,2) {D};
    \node[shape=circle,draw=black,line width=0.25mm] (E) at (1.2,3) {E};
    \node[align=left] at (0.9,4) {\textbf{Reductions}};
    \path [->,line width=0.25mm](A) edge node[left] {} (B);
    \path [->,line width=0.25mm](B) edge node[left] {} (D);
    \path [->,line width=0.25mm](C) edge node[left] {} (D);
    \path [->,line width=0.25mm](D) edge node[left] {} (E);
    \node[align=left] at (4.5,0) {$n$};
    \node[align=left] at (4.5,1) {$n,k$};
    \node[align=left] at (4.5,2) {$n,k_{min},k_{max}$};
    \node[align=left] at (4.5,3) {$n,K$};
    \node[align=left] at (4.5,4) {\textbf{Parameters}};
    \node[align=left] at (9,0) {without restriction};
    \node[align=left] at (9,1) {at most $k$ -- exactly $k$};
    \node[align=left] at (9,2) {between $k_{min}$ and $k_{max}$};
    \node[align=left] at (9,3) {any $k \in K$};
    \node[align=left] at (9,4) {\textbf{Behavior}};
    \end{tikzpicture}
    \caption{Overview of the set partition enumeration problems. The figure shows the reductions among Problems A, B, C, D and E, their parameters and their behavior. An edge $U \to V$ shows that problem $U$ can be reduced to problem $V$ with parameter adjustment. Problem E is the most general problem discussed in this document.}
    \label{fig:reductions}
\end{figure}
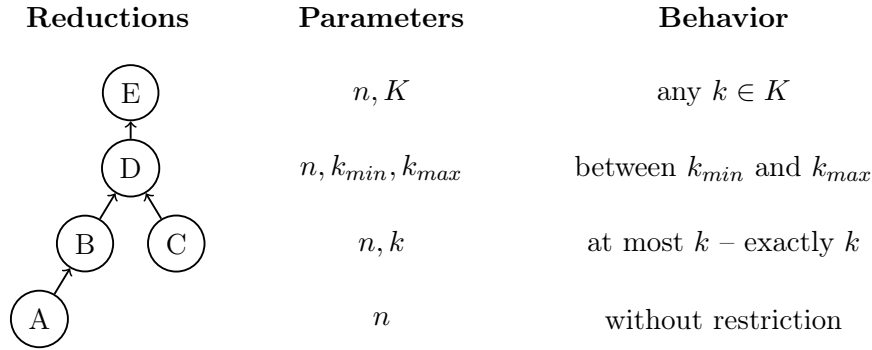

It is immediately evident that the problem of enumerating all partitions of a set is a special case of another problem: enumerating all partitions of a set with at most $k$ blocks; if $k=n$, then the two problems are equivalent. Conversely, the second problem is a special case of a third problem: enumerating all partitions of a set with the number of blocks residing between $k_{min}$ and $k_{max}$, from which it is reduced for $k_{min}=1$ and $k_{max}=k$. The latter problem is also a more general case of enumerating all partitions with exactly $k$ blocks. The process can be further generalized by the problem of enumerating partitions of $K=\{k_1,k_2,k_3,\dots\}$ blocks, where $k_i$ are arbitrary integer block counts greater than zero. More formally, we define 5 lexicographic set partition enumeration problems:
\begin{description}
    \item [Problem A] Enumerate all partitions of a set.
    \item [Problem B] Enumerate all partitions of a set with at most $k$ blocks.
    \item [Problem C] Enumerate all partitions of a set with exactly $k$ blocks.
    \item [Problem D] Enumerate all partitions of a set with the number of blocks between $k_{min}$ and $k_{max}$.
    \item [Problem E] Enumerate all partitions of a set with the number of blocks in the $K=\{k_1,k_2,k_3,\dots\}$.
\end{description}
These problems are stated, and later studied, in this order, as it appears to be a natural sequence of the algorithms involved. An overview of these problems can be seen in Figure~\ref{fig:reductions}.

Knuth describes a classic set partition enumeration algorithm which is due to \citet{10.1145/367651.367661} and enumerates all partitions of a set in lexicographic order (Problem A). Other authors have since improved Hutchinson's work and incrementally achieved better running times \citep{nijenhuis2014combinatorial,110002673365,10.1093/comjnl/31.3.283,10.1093/comjnl/32.3.281}. These lexicographic enumeration algorithms can be iterative or recursive and are often based on the discovery tree formed by increasing number of elements $n$. In this document, we additionally articulate Problems B, C, D and E and adopt more specialized algorithms for the solution of these problems. The algorithms presented are formulated without \textit{jump} logic (e.g., \textit{goto}, \textit{break}) or recursive structures, as these are often difficult to understand and can lead to error-prone code. They abide by modern development practices, are easy to implement and, if implemented naturally, don't lead to write-only code.


For this reason, we reduce the definition of each algorithm into the definition of two components:
\begin{enumerate}
    \item The first partition that is returned by the enumerator.
    \item The function \texttt{next(p,s)} which determines the next partition that is visited, given the previous partition \texttt{p} and any auxiliary state \texttt{s} that the algorithm uses.
\end{enumerate}
Using these components, a full lexicographic enumeration is possible via an iterative approach, where in each iteration a new restricted growth string is generated. We prove that all presented algorithms use extra space $\Theta(n)$ and the \texttt{next} method runs in constant amortized time. We use the restricted growth string notation for the algorithms as it is easier to understand and can be implemented simply with an integer array.


This document is inspired by the work of \citet{orlov2002efficient}, which is aimed at programmers undertaking the implementation of partition enumeration algorithms. Additionally, we describe the problems and their solutions in a learner's perspective as well, with formal complexity analysis. Finally, we introduce the additional problems B, D and E that are not mentioned in that report, and form the generalizations of the simpler problems A and C.



This document is structured as follows: Section~\ref{sec:preliminaries} is a short preliminaries overview and the Problems A, B, C, D and E are presented individually in Sections~\ref{sec:problem-a} through~\ref{sec:problem-e}.

\section{Preliminaries}
\label{sec:preliminaries}

It has been previously shown that the number of possible ways that a set of $n$ elements can be partitioned is defined by the $n$-th \textit{Bell number}, denoted by $\mathcal{B}_n$. Likewise, $\mathcal{S}(n,k)$ is the \textit{Stirling number of the second kind} and represents the number of ways to partition a set of $n$ objects into $k$ non-empty blocks.

Bell numbers and Stirling numbers of the second kind are related via the following equation:
\begin{equation}
    \mathcal{B}_n = \sum_{k=0}^n \mathcal{S}(n,k).
\end{equation}

We also note the following asymptotic relations that will be used for the analysis of the algorithms in the following sections:
\begin{equation}
    \label{eq:asymptotic}
    \frac{\mathcal{B}_{n+1}}{\mathcal{B}_n} \sim \frac{n}{\ln n} \text{~~~and~~~} \mathcal{S}(n,k) \sim \frac{k^n}{k!}.
\end{equation}
The notation we use throughout this document to describe asymptotic equivalence is defined in~\citet[Section 1.4]{de1981asymptotic}. In particular, two functions $f(x)$ and $g(x)$ are asymptotically equivalent as $x \to \infty$ if the quotient $f(x)/g(x)$ tends to unity and we note this relation as $f(x) \sim g(x)$.

Finally, we define $\mathcal{S}^+(n,k)$ as
\begin{equation}
    \mathcal{S}^+(n,k) = \sum_{i=0}^k \mathcal{S}(n,i).
\end{equation}
Regarding the quantity $\mathcal{S}^+(n+1,k)/\mathcal{S}^+(n,k)$, because $\mathcal{S}(n,i) = i^n/i!$, the limit of the fraction of sums (asymptotically) only depends on the most dominant factor, which is $k \cdot \mathcal{S}(n,k) = k^{n+1}/k!$ for the numerator and $\mathcal{S}(n,k) = k^n/k!$ for the denominator. Hence, it also holds that asymptotically
\begin{equation}
    \label{eq:asymptotic-2}
    \frac{\mathcal{S}^+(n+1,k)}{\mathcal{S}^+(n,k)} \sim \frac{\mathcal{S}(n+1,k)}{\mathcal{S}(n,k)} \sim k.
\end{equation}

\section{Problem A}
\label{sec:problem-a}

Problem A is the most basic problem of the set partition enumeration problem and can be solved using Hutchinson's algorithm. In this section, we present a modified version of this algorithm that is easier to be implemented in modern, high-level programming languages. The partitions are returned in the lexicographic order of their respective restricted growth strings. For example, for $n=4$, \texttt{next(0101) = 0102} and \texttt{next(0112) = 0120}. The full lexicographic order of the $n=4$ partitions were shown earlier in Section~\ref{sec:introduction}.

Similar to the original algorithm, the modification uses auxiliary state array $b = [b_1, b_2, \dots, b_n]$, where $b_i = \text{max}(a_1, a_2, \dots, a_{i-1})$. If adjusted successively, then $b_i = \text{max}(a_{i-1}, b_{i-1})$. Hence, the restrictions of Equation~\ref{eq:restricted-growth-string} can be simplified to
\begin{equation}\label{eq:conditions-algorithm-a}
    a_1 = 0 \text{~~~and~~~} a_{j} \le 1 + b_{j} \text{~for~} 1 \le j < n.
\end{equation}
The arrays $a$ and $b$ are initialized with zeroes. Starting from position $n$ of the $a$ array and moving towards position $1$, we find the right-most digit of the restricted growth string that can be incremented without violating the conditions of Equation~\ref{eq:conditions-algorithm-a}. The one-based index $c$ that will be incremented is, thus, the first to satisfy both $a_c < n - 1$ and $a_c \le b_c$, otherwise after the increment we would have an invalid partition. After finding such index $c$, we move in the opposite direction from $c+1$ to $n$ and zero out the digits (since we have performed an increment on a more significant digit) while adjusting the values of $b_i = \text{max}(a_{i-1}, b_{i-1})$. The process is repeated until no index can be incremented (except the first). More formally, the \texttt{next} method of this algorithm is described by the following steps (one-based indices notation).

\begin{description}
    \item [Algorithm V] Accepts the array $a$ representing the previous partition and the auxiliary state variable $b$ representing the prefix maxima of array $a$. Changes the array $a$ in-place to point to the next partition and updates the array $b$ to reflect the changes.
    \item [V1] Set [$c \gets n$].
    \item [V2] While [$a_c = n - 1$ Or $a_c > b_c$] Do [$c \gets c - 1$].
    \item [V3] If [$c = 1$] Report the end of enumeration.
    \item [V4] Set [$a_c \gets a_c + 1$].
    \item [V5] For [$i$] From [$c + 1$] To [$n$], execute the following steps:
    \begin{description}
        \item [V5.1] Set [$a_i \gets 0$].
        \item [V5.2] Set [$b_i \gets \text{max}(a_{i-1},b_{i-1})$].~~~$\blacksquare$
    \end{description}
\end{description}

The variable $a$ can be implemented using either an array or a linked list and there is theoretically no benefit in using either of these data structures since random access is not required. In practice, however, arrays are preferable due to their ubiquity, smaller overhead and cache efficiency. The same arguments can be made for the $b$ array.

We estimate the complexity of Algorithm V via a recurrence formula. Assume $f(n)$ is the average number of steps (array changes) involved in a full enumeration of all partitions, which is equal to the total number of steps over $\mathcal{B}_n$. The total number of steps is equal to the number of advances where only the least significant digit changes plus the complexity of the rest of the cases, which can be reduced to the complexity of the same problem with $n-1$ elements where all steps are increased by 1 (because the least significant digit has to be traversed). The cases of the latter scenario are then $\mathcal{B}_{n-1}$, via which we can also say that the number of least significant digit advances are $\mathcal{B}_n - \mathcal{B}_{n-1}$ because both cases need to sum to $\mathcal{B}_n$. Thus,
\begin{align*}
    f(n) &= \frac{\mathcal{B}_n - \mathcal{B}_{n-1} + \mathcal{B}_{n-1} (f(n-1)+1)}{\mathcal{B}_n} \\
         &= 1 - \frac{\mathcal{B}_{n-1}}{\mathcal{B}_{n}} + \frac{\mathcal{B}_{n-1}}{\mathcal{B}_{n}} f(n-1) + \frac{\mathcal{B}_{n-1}}{\mathcal{B}_{n}}.
\end{align*}
Considering Equation~\ref{eq:asymptotic}, $f(n)$ can asymptotically be written as
\begin{equation*}
    f(n) \sim 1 + \frac{\ln n}{n} f(n-1)
\end{equation*}
and it can easily be shown that $f(n)$ is decreasing and $\lim_{n \to \infty} f(n) = 1$. Hence, the amortized number of primitive array operations are $\Theta(1)$. Another perspective of this result is that the probability of the \texttt{next} method returning after a single change to the least significant digit is $1-\ln n / n$, which increases with $n$ and tends towards 1. Hence, it becomes increasingly likely that the \texttt{next} method will only have to perform a single change to the $a$ array. The extra space used is the $b$ array which corresponds to $\Theta(n)$ space requirement. It can be easily shown that the worst case complexity of the \texttt{next} method is linear in respect to $n$ and occurs for $c=1$, in which case the $a$ array will be traversed twice.

\section{Problem B}
\label{sec:problem-b}

Problem B consists in enumerating the partitions of a set with at most $k$ blocks. As a result, Problem A can reduce to Problem B if we set $k=n$. However, the enumerations of problem B are a subset of those of Problem A and, thus, it would appear that Problem B can simply be solved by using Algorithm V and filtering only the $k$-partitions. Such filtering can be performed in constant time; the quantity $\text{max}(a_n,b_n)$ shows how many blocks the current partition has. However, despite the filtering process being constant, this algorithm would still not have constant amortized advance, as shown in the following paragraph.

Suppose such algorithm exists and solves Problem B. Then, the iterations that would have to be performed are $\mathcal{B}_n$, and the enumerations would be $\mathcal{S}^+(n,k)$. Thus, the amortized number of steps required to advance the enumeration is
\begin{equation*}
    \mathcal{T}_k(n) = \frac{\mathcal{B}_n}{\mathcal{S}^+(n,k)}.
\end{equation*}
Hence, due to Equations~\ref{eq:asymptotic} and~\ref{eq:asymptotic-2}
\begin{align*}
    \frac{\mathcal{T}_k(n+1)}{\mathcal{T}_k(n)} &=\frac{\mathcal{B}_{n+1}}{\mathcal{B}_n} \cdot  \frac{\mathcal{S}^+(n,k)}{\mathcal{S}^+(n+1,k)} \\
      &\sim \frac{n}{\ln n} \cdot \frac{1}{k},
\end{align*}
which is increasing with $n$ and tends towards infinity. Hence, $\mathcal{T}_k(n)$ cannot represent constant time advance as in the latter case the fraction should tend to a constant. As a result, Problem B cannot be solved in constant time via filtering the partitions of Algorithm V.

Instead, we perform a simple modification to Algorithm V that guarantees that there will never be any partition with more than $k$ blocks. In the terminology of the restricted growth strings, the constraint can be imprinted as the absence of any number that is greater or equal to $k$. Such algorithm is very simple to design with a small modification on step V2:

\begin{description}
    \item [Algorithm W] Accepts the array $a$ representing the previous partition, the setting $k$ that represents the \textit{maximum} number of blocks in any partition, and the auxiliary state variable $b$ representing the prefix maxima of array $a$. Changes the array $a$ in-place to point to the next partition and updates the array $b$ to reflect the changes. No change is performed on $k$.
    \item [W1] Set [$c \gets n$].
    \item [W2] While [$a_c = k-1$ Or $a_c > b_c$] Do [$c \gets c - 1$].
    \item [W3] If [$c = 1$] Report the end of enumeration.
    \item [W4] Set [$a_c \gets a_c + 1$].
    \item [W5] For [$i$] From [$c + 1$] Up To [$n$], execute the following steps:
    \begin{description}
        \item [W5.1] Set [$a_i \gets 0$].
        \item [W5.2] Set [$b_i \gets \text{max}(a_{i-1},b_{i-1})$].~~~$\blacksquare$
    \end{description}
\end{description}

The only step that has changed is V2 where the condition $a_c = n-1$ is transformed into $a_c = k-1$. This change guarantees that the maximum number of blocks on any partition will be $k$, while maintaining the lexicographic order of the partitions.

Although Algorithm W is only a slight modification of Algorithm V, the complexity analysis differs. The overall number of iterations is $\mathcal{S}^+(n,k)$ and in each one of them, we perform $n-c$ steps. The probability of performing more than 1 step is $\mathcal{S}^+(n-1,k)/\mathcal{S}^+(n,k)$. Thus, for the average number of steps $f(n)$ it holds that
\begin{align*}
    f(n) &= \frac{ \mathcal{S}^+(n,k) - \mathcal{S}^+(n-1,k) + \mathcal{S}^+(n-1,k) \cdot (f(n-1) + 1) }{\mathcal{S}^+(n,k)} \\
         &= 1 - \frac{1}{k} + \frac{1}{k} \cdot f(n-1) + \frac{1}{k} = 1 + \frac{1}{k} \cdot f(n-1) \\
         &= 1 + \frac{1}{k} + \frac{1}{k^2} + \dots + \frac{1}{k^n} \sim 1 + \frac{1}{k-1},
\end{align*}
which is independent of $n$ and, as a result, dictates a $\Theta(1)$ amortized advance. An interesting observation is that the average number of array operations of Algorithm W tends towards $1+1/(k-1)$ which, unlike Algorithm V, is higher than 1 for finite $k$. Experimental observations suggest that the number of steps converge rapidly, even for $n<15$. If $k=n$, then the number of steps performed by Algorithm W tends towards 1 and reduces to Algorithm V. Hence, Algorithm W becomes faster (in respect to the average \texttt{next} call) as $k$ increases, with the amortized number of steps per call to the \texttt{next} method ranging from 1 in the best case to 2 in the worst case ($k=2$). The extra memory of Algorithm W is $\Theta(n)$ due to the $b$ array. Similar to Algorithm V, the worst case complexity of the \texttt{next} method is also linear in respect to $n$ as the $a$ array is traversed twice.

\section{Problem C}
\label{sec:problem-c}

Problem C consists in enumerating the partitions of a set with exactly $k$ blocks. Similar to Algorithm W, an algorithm for Problem C would be inefficient by a simple filtering of Algorithm V, despite the enumerations of Problem C being a subset of the enumerations of Problem A. The iterations that would have to be performed are $\mathcal{B}_n$, and the enumerations would be $\mathcal{S}(n,k)$. Thus, the amortized number of steps required to advance the enumeration is
\begin{equation*}
    \mathcal{T}_k(n) = \frac{\mathcal{B}_n}{\mathcal{S}(n,k)}.
\end{equation*}
Hence, due to Equation~\ref{eq:asymptotic}
\begin{align*}
    \frac{\mathcal{T}_k(n+1)}{\mathcal{T}_k(n)} &=\frac{\mathcal{B}_{n+1}}{\mathcal{B}_n} \cdot  \frac{\mathcal{S}(n,k)}{\mathcal{S}(n+1,k)} \\
      &\sim \frac{n}{\ln n} \cdot \frac{k^n}{k^{n+1}} = \frac{1}{k} \cdot \frac{n}{\ln n},
\end{align*}
which is increasing with $n$ and tends towards infinity. Hence, $\mathcal{T}_k(n)$ cannot represent constant time advance.

A constant time filtering via Algorithm W, however, appears to be possible; this property will allow us to design a constant time algorithm for Problem C. Consider, for example, a modification to Algorithm W that filters only the $k$-partitions (i.e., only those that have the number $k$ in the restricted growth string). The filtering operation is trivial to design in constant time by leveraging the maximum value of the $a$ array. Hence, the iterations of this algorithm would be $\mathcal{S}^+(n,k)$ and the enumerations of Problem C are $\mathcal{S}(n,k)$. As a result, the amortized number of steps required to advance the enumeration is
\begin{equation*}
    \mathcal{T}_k(n) = \frac{\mathcal{S}^+(n,k)}{\mathcal{S}(n,k)}.
\end{equation*}
It can be shown that $\mathcal{T}_k(n)$ is asymptotically a constant time operation by considering that
\begin{equation*}
    \frac{\mathcal{S}(n,k)}{\mathcal{S}(n,k-1)} = \left( \frac{k}{k-1} \right) ^ n \cdot \frac{1}{k},
\end{equation*}
which approaches infinity as $n \to \infty $. Thus, $\mathcal{S}(n,k)$ becomes infinitely bigger than $\mathcal{S}(n,k-1)$ and, therefore, all the mass of the $\mathcal{S}^+(n,k)$ sum is being concentrated to its last term $\mathcal{S}(n,k)$. Therefore, for a fixed $k$:
\begin{equation}\label{eq:limit-striling-sum}
    \mathcal{S}^+(n,k) \sim \mathcal{S}(n,k),
\end{equation}
which leads to $\mathcal{T}_k(n) = \Theta(1)$. This process implies the existence of Algorithm X which can be developed by simply filtering the partitions returned by Algorithm W and asymptotically perform $1+1/(k-1)$ operations and an extra two operations for the filtering process: the access of $a_n$ and $b_n$. Naturally, Algorithm X also enumerates the partitions in their lexicographic order.

\begin{description}
    \item [Algorithm X] Accepts the array $a$ representing the previous partition, the setting $k$ that represents the \textit{exact} number of blocks in any partition, and the auxiliary state variable $b$ representing the prefix maxima of array $a$. Changes the array $a$ in-place to point to the next partition and updates the array $b$ to reflect the changes. No change is performed on $k$.
    \item [X1] Set [$m \gets 0$].
    \item [X2] Execute the following statements While [$m \ne k$].
    \begin{description}
        \item [X2.1] Invoke Algorithm W.
        \item [X2.2] Set [$m \gets \text{max}(a_n, b_n)$].~~~$\blacksquare$
    \end{description}
\end{description}

The algorithm includes the $m$ variable that keeps track of the maximum number in the $a$ array which corresponds to the number of blocks in the next partition. The initial value of the $a$ array is $0^{n-k} 01 \dots (k-1)$ as this is the first lexicographic $k$-partition. As proven earlier, the amortized running time of an advance is $\Theta(1)$ and it uses extra $\Theta(n)$ memory. From the arguments leading to Equation~\ref{eq:limit-striling-sum}, it is implied that, asymptotically, step X2 will only get executed once and, as a result, the worst case complexity of Algorithm X is the same as Algorithm W, which is linear in respect to $n$.

Algorithm X achieves a constant time advance via an extension to Algorithm W that skips the invalid partitions one at a time. A natural extension to Algorithm X can easily be stated by observing the pattern of partitions returned by Algorithm W. Below is a small contiguous subset for $n=7,k=3$ where the bold partitions indicate the (valid for Problem C) $3$-partitions.
\begin{quote}
    \centering
    \textbf{100002} 0100010 0100011 \textbf{0100012} \textbf{0100020} \textbf{0100021} \textbf{0100022} 0100100 0100101 \textbf{0100102} 0100110 0100111 \textbf{0100112} \textbf{0100120} \textbf{0100121} \textbf{0100122} \textbf{0100200} \textbf{0100201} \textbf{0100202} \textbf{0100210} \textbf{0100211} \textbf{0100212} \textbf{0100220} \textbf{0100221} \textbf{0100222} 0101000 0101001 \textbf{0101002} 0101010 0101011 \textbf{0101012} 0101020 \textbf{0101021} \textbf{0101022} 0101100 0101101 \textbf{0101102} 0101110 0101111 \textbf{0101112} \textbf{0101120} \textbf{0101121} \textbf{0101122} \textbf{0101200} \textbf{0101201} \textbf{0101202}
\end{quote}
The valid partitions are not randomly distributed but instead usually come in sequences. In fact, the enumerations can be reduced to alternating sequences of valid and invalid partitions. This phenomenon is more obvious for larger values of $k$ where there are longer invalid sequences. Careful inspection of the enumerations reveals that all invalid sequences begin with a partition for which $a_n=0$, because the lexicographically previous partition has $a_n=k-1$ and is, thus, valid. Based on this observation, we define Algorithm Y as another modification of Algorithm W that skips sequences of invalid partitions instead of determining whether to visit the partitions one by one.

Thus, the problem reduces to finding the next valid partition, given any invalid partition. For example, for the sequence above, the next valid partition from $0100010$ is $0100012$. The way to achieve this is to move a pointer $i$ from right to left (from LSD to MSD) and set $a_i$ as the highest number, lower than $k$, that is not in the array, until this value is found in the array. This process is depicted in the description below.

\begin{description}
    \item [Algorithm Y] Accepts the array $a$ representing the previous partition, the setting $k$ that represents the \textit{exact} number of blocks in any partition, and the auxiliary state variable $b$ representing the prefix maxima of array $a$. Changes the array $a$ in-place to point to the next partition and updates the array $b$ to reflect the changes. No change is performed on $k$.
    \item [Y1] Execute Algorithm W.
    \item [Y2] If [$\text{max}(a_n,b_n) \ne k-1$]:
    \begin{description}
        \item [Y2.1] For [$(i,k_0)$] From [$(n,k-1)$] Down To [$(1,1)$] As Long As [$k_0 > b_i$].
        \begin{description}
            \item [Y2.1.1] Set [$a_i \gets k_0$].
            \item [Y2.1.2] Set [$b_i \gets k_0 - 1$].~~~$\blacksquare$
        \end{description}
    \end{description}
\end{description}

The initial value of Algorithm Y is the same is Algorithm X.

Algorithm Y performs as many steps as Algorithm W because it is easy to show that step Y2.1 will asymptotically not get executed. Specifically, step Y2.1 will be executed only if the generated partition is not a $k$-partition. The probability of a partition which is at most a $k$-partition being exactly a $k$-partition is asymptotically
\begin{equation*}
    p = \lim_{n \to \infty} \frac{\mathcal{S}(n,k)}{\mathcal{S}^+(n,k)} = 1
\end{equation*}
because of Equation~\ref{eq:limit-striling-sum}. Thus, asymptotically, the frequency at which step Y2.1 is executed is approaching 0 and, as a result, Algorithm Y performs as many steps as Algorithm W plus the comparison step in Y2. Similar to algorithm X, the worst case complexity of Algorithm Y is also linear in respect to $n$ since step Y2.1 will asymptotically not get executed.

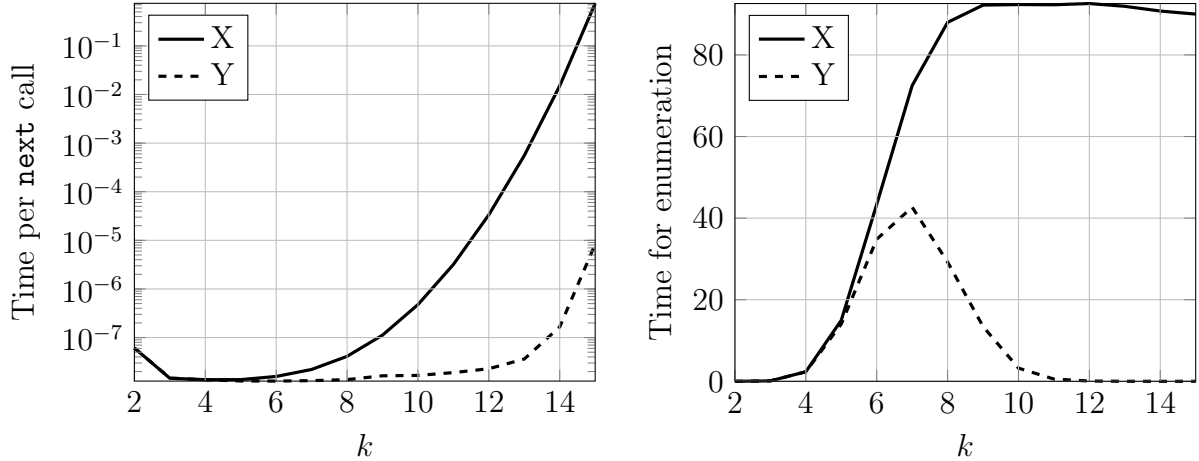
\begin{figure}
    \centering
    \begin{tikzpicture}
      \begin{axis}[
        no markers,
        legend pos=north west,
        xlabel=$k$, ylabel=Time per \texttt{next} call,
        width=0.48\textwidth,
        ymode=log,
        enlargelimits=false, clip=false, axis on top,
        grid = major
      ]
      \addplot[very thick,black] table[x=k,y=Xpn] {comparison.csv};
      \addplot[very thick,black,dashed] table[x=k,y=Ypn] {comparison.csv};
      \legend{X,Y}
      \end{axis}
    \end{tikzpicture}
    \quad
    \begin{tikzpicture}
      \begin{axis}[
        no markers,
        legend pos=north west,
        xlabel=$k$, ylabel=Time for enumeration,
        width=0.48\textwidth,
        enlargelimits=false, clip=false, axis on top,
        grid = major
      ]
      \addplot[very thick,black] table[x=k,y=X] {comparison.csv};
      \addplot[very thick,black,dashed] table[x=k,y=Y] {comparison.csv};
      \legend{X,Y}
      \end{axis}
    \end{tikzpicture}
    \caption{Amortized time in seconds per \texttt{next} call (left) and full enumeration (right) for Algorithms X and Y measured experimentally. The left y-axis is logarithmic for aesthetic reasons as the curve is quite steep for large $k$ in linear scale.}
    \label{fig:comparison}
\end{figure}

Experimental observations suggest that Algorithm Y is much faster than Algorithm X. Figure~\ref{fig:comparison} shows plots of their execution times in seconds for $n=16$ and values of $k$ ranging from 2 to 15. The left plot shows the amortized time per call to the \texttt{next} function while the plot to the right shows the time in seconds for the complete enumeration. The differences scale profoundly as $k$ increases and this is because the number of skips that Algorithm Y takes advantage of are dramatically increased as $k$ increases. In other terms, Algorithm X performs operations proportional to $\mathcal{S}^+(n,k)$, which is the amount of times that Algorithm W has to be invoked, while Algorithm Y only performs operations proportional to $\mathcal{S}(n,k)$. The difference between these two quantities increases with $k$ because Stirling numbers of the second kind are decreasing after their peak for specific $n$ (which is $k=7$ in this case), despite both algorithms running in constant amortized time. For this reason, we recommend Algorithm Y over Algorithm X despite them both having constant complexity.

\section{Problem D}
\label{sec:problem-d}

Problem D is the generalization of problems A, B and C. It consists in enumerating the partitions of a set, whose number of blocks are between $k_{min}$ and $k_{max}$.

For this problem, we define a new algorithm, Algorithm Z, that builds upon the concept of Algorithm Y where we skip the whole invalid sequence before generating the next partition. Instead of backtracking, however, after an invalid partition has been generated (Y2.1), Algorithm Z has a return process on par with step W5 which is split into two parts and, thus, no backtracking is necessary. The return process of Algorithm Z aims at replacing the subarray $a_{c+1} a_{c+2} \dots a_n$ with the lexicographically smallest string that guarantees a $k_{min}$-partition. The first part of the return process consists in filling the longest possible sequence with zeroes (Z7). For the second part of the return, Algorithm Z fills the remaining places so that $a_n=k_{min}-1$, $a_{n-1}=k_{min}-2$ or simply $a_{n-i}=k_{min}-i-1$ (Z8).

\begin{description}
    \item [Algorithm Z] Accepts the array $a$ representing the previous partition, the setting $k_{min}$ that represents the \textit{minimum} number of blocks in any partition, the setting $k_{max}$ that represents the \textit{maximum} number of blocks in any partition, and the auxiliary state variable $b$ representing the prefix maxima of array $a$. Changes the array $a$ in-place to point to the next partition and updates the array $b$ to reflect the changes. No change is performed on $k_{min}$ or $k_{max}$.
    \item [Z1] Set [$c \gets n$].
    \item [Z2] While [$a_c = k_{max} - 1$ Or $a_c > b_c$] Do [$c \gets c - 1$].
    \item [Z3] If [$c = 1$] Report the end of enumeration.
    \item [Z4] Set [$a_c \gets a_c + 1$].
    \item [Z5] Set [$b_{c+1} \gets \text{max}(a_c, b_c)$].
    \item [Z6] Set [$z \gets b_{c+1} + n - c - k_{min}$].
    \item [Z7] Execute [$z$] times and as long as $c \le n$:
    \begin{description}
        \item [Z7.1] Set [$a_c \gets 0$].
        \item [Z7.2] Set [$b_{c+1} \gets b_c$].
        \item [Z7.3] Set [$c \gets c + 1$].
    \end{description}
    \item [Z8] For [$i$] From [$c$] To [$n$], execute the following steps:
    \begin{description}
        \item [Z8.1] Set [$a_i \gets b_i + 1$].
        \item [Z8.2] Set [$b_{i+1} \gets b_i$].~~~$\blacksquare$
    \end{description}
\end{description}

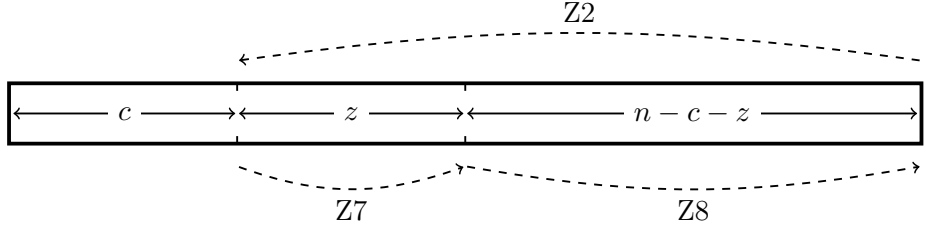
\begin{figure}
    \centering
    \small
    \begin{tikzpicture}
    \draw [line width=0.5mm] (-6,-0.4) -- (-6,0.4) -- (6,0.4) -- (6,-0.4) -- cycle;
    \draw [line width=0.25mm] (-3,0.4) -- (-3,0.3);
    \draw [line width=0.25mm] (-3,-0.4) -- (-3,-0.3);
    \draw [line width=0.25mm] (0,0.4) -- (0,0.3);
    \draw [line width=0.25mm] (0,-0.4) -- (0,-0.3);
    \node [inner sep=0cm] (A) at (6,0.7) {};
    \node [inner sep=0cm] (B) at (-3,0.7) {};
    \node [inner sep=0cm] (C) at (-3,-0.7) {};
    \node [inner sep=0cm] (D) at (0,-0.7) {};
    \node [inner sep=0cm] (E) at (6,-0.7) {};
    \node [inner sep=0.5mm] (F) at (-6,0) {};
    \node [inner sep=0cm] (G) at (-3,0) {};
    \node [inner sep=0cm] (H) at (0,0) {};
    \node [inner sep=0.5mm] (I) at (6,0) {};
    \path [<->,line width=0.25mm] (F) edge [] node [fill=white] {$c$} (G);
    \path [<->,line width=0.25mm] (G) edge [] node [fill=white] {$z$} (H);
    \path [<->,line width=0.25mm] (H) edge [] node [fill=white] {$n-c-z$} (I);
    \path [->,line width=0.25mm] (A) edge [bend right=8, above, dashed] node {Z2} (B);
    \path [->,line width=0.25mm] (C) edge [bend right=20, below, dashed] node {Z7} (D);
    \path [->,line width=0.25mm] (D) edge [bend right=10, below, dashed] node {Z8} (E);
    \end{tikzpicture}
    \caption{The operation of Algorithm Z. Initially, step Z2 searches for the largest incrementable index $c$. Direction is then switched to continue with step Z7 that fills the region with $z$ zeroes. During the final step Z8, the algorithm fills the remaining places with the lexicographically minimum string that guarantees a $k_{min}$-partition.}
    \label{fig:algorithm-z}
\end{figure}

The algorithm includes an extra index in the $b$ array for convenience; it's not utilized but it prevents unnatural \textit{if} statements during steps Z7.2 and Z8.2 to check for array boundaries. Steps Z1-Z5 are performed in the same way as Algorithm W as the algorithm seeks the right-most element that can be incremented. During step Z7, the $z$ digits that follow are set to 0 after the $z$ value is being calculated in step Z6. This value is computed by considering what the value of $a_n$ would be if we kept increasing the $a_c$ values until the end of the array. If this value is higher than $k_{min}$, it means that we can still put an extra 0, as our goal, in the worst case, is to terminate with $a_n=k_{min}-1$. If $z$ is zero, step Z7 is skipped and step Z8 is executed immediately. The value $z$ cannot be negative as that would mean that creating a partition that has over $k_{min}$ number of blocks is impossible, which contradicts the fact that the previous partition was at least a $k_{min}$-partition. Finally, during step Z8 we fill up the rest of the array so that $a_i = k_{min}-i-1$. In conclusion, steps Z7 and Z8 guarantee that the next partition formed by the digits $c+1$ through $n$ will be the lexicographically smallest that guarantees a $k_{min}$-partition. The initial value of the $a$ array is $0^{n-k_{min}} 012 \dots (k_{min}-1)$ as this is the first lexicographic $k_{min}$-partition. A diagram of the operation of Algorithm Z is shown in Figure~\ref{fig:algorithm-z}.

An interesting property of Algorithm Z is that, in certain cases, step Z8 is not required. Skipping step Z8 will result in a restricted growth string that doesn't respect Equation~\ref{eq:restricted-growth-string} and, as a result, it would appear that such a modification would result in wrong enumeration. This is, however, not always the case as Algorithm Z never returns isomorphic partitions, even with the absence of step Z8, for example it cannot return both 00100023 and 00100067. Thus, when the growth string is transformed into a partition (for example an object of a \texttt{Partition} class), these two growth strings are likely to be pointing to equal objects; this property also depends on the actual transformation implementation. Although this property, if leveraged, can lead to faster code, it is not recommended to do so as it will give rise to new issues and the benefits are often outweighed by complications, for example the possible unchecked increase of values in the $a$ array.

Algorithm Z can be analyzed in the same way as the rest of the algorithms. For convenience, we define
\begin{equation*}
    \mathcal{Q}(n,u,v) = \sum_{k=u}^v \mathcal{S}(n,k).
\end{equation*}
Then the probability of performing more than one step is
\begin{equation*}
    q = \frac{\mathcal{Q}(n-1,k_{min},k_{max})}{\mathcal{Q}(n,k_{min},k_{max})} = \frac{\mathcal{S}(n-1,k_{max})}{\mathcal{S}(n,k_{max})} = \frac{1}{k_{max}}
\end{equation*}
for the same arguments that allowed us to formulate Equation~\ref{eq:asymptotic-2} and Equation~\ref{eq:limit-striling-sum}. Given that $k_{min}$ does not have an impact on the probability $q$ we might remove it:
\begin{equation*}
    q = \frac{\mathcal{Q}(n-1,k_{max})}{\mathcal{Q}(n,k_{max})} = \frac{1}{k_{max}}.
\end{equation*}
Thus,
\begin{align*}
    f(n) &= \frac{\mathcal{Q}(n,k_{max})-\mathcal{Q}(n-1,k_{max})+\mathcal{Q}(n-1,k_{max}) \cdot (f(n-1)+1)}{\mathcal{Q}(n,k_{max})} \\
         &= 1 - \frac{1}{k_{max}} + \frac{1}{k_{max}} \cdot f(n-1) + \frac{1}{k_{max}} = 1 + \frac{1}{k_{max}} \cdot f(n-1).
\end{align*}
This is the same result as Algorithm W. Thus, the average number of operations that Algorithm Z performs are proportional to
\begin{equation*}
    1 + \frac{1}{k_{max}-1},
\end{equation*}
which is independent of $n$ and implies that algorithm Z runs in constant amortized time. Initially, we removed $k_{min}$ because it was inconsequential, considering the reasoning of Equation~\ref{eq:asymptotic-2} and Equation~\ref{eq:limit-striling-sum}: in a sum of contiguous Stirling numbers of the second kind, only the highest term is relevant in an asymptotic perspective. Another way to look at the absence of $k_{min}$ is because as $n$ increases, the percentage of partitions with smaller number of blocks become less common. For example for $k_{min}=2$ and $k_{max}=3$, the percentages of 2-partitions in the enumeration as $n$ goes from 3 to 10 are 0.75, 0.54, 0.38, 0.26, 0.17, 0.12, 0.08, 0.05. Similar to the previous algorithms, the worst case complexity of Algorithm Z occurs when $c$ is lowest, in which case the $a$ array is traversed twice.

Before we close this section, we would like to make a short mention to the problem of lexicographically enumerating the partitions of a set in reverse order. Sometimes the requirements of a problem dictate the enumeration be executed in reverse order; such operation can easily be produced from the existing algorithms mentioned in this work. As an example, Algorithm Z can be transformed to an algorithm that enumerates the partitions in reverse lexicographic order by inverting its logic. Specifically, during step Z2, instead of searching for the first index that can be incremented, we will search for the first index that can be decremented; this is the first index that is not 0 and at the same time, after its reduction, there is enough ``space'' to reach $k_{min}$ at the end of the string. During the respective steps Z7 and Z8, we create the lexicographically maximum string by increasing the values of $a_n$ until $k_{max}-1$ has been reached, and fill the remaining places with $k_{max}-1$. The reverse version of Algorithm Z is formulated below.

\begin{description}
    \item [Algorithm Z*] Accepts the array $a$ representing the next partition, the setting $k_{min}$ that represents the \textit{minimum} number of blocks in any partition, the setting $k_{max}$ that represents the \textit{maximum} number of blocks in any partition, and the auxiliary state variable $b$ representing the prefix maxima of array $a$. Changes the array $a$ in-place to point to the previous partition and updates the array $b$ to reflect the changes. No change is performed on $k_{min}$ or $k_{max}$.
    \item [Z*1] Set [$c \gets n$].
    \item [Z*2] While [($c > 1$) And ($a_c = 0$ Or $k_{min} - b_i > n - i$)] Do [$c \gets c - 1$].
    \item [Z*3] If [$c = 1$] Report the end of enumeration.
    \item [Z*4] Set [$a_c \gets a_c - 1$].
    \item [Z*5] Set [$b_{c+1} \gets \text{max}(a_c, b_c)$].
    \item [Z*6] For [$i$] From [$c+1$] To [$n$] As Long As [$b_i < k_{max}-1$]:
    \begin{description}
        \item [Z*6.1] Set [$a_i \gets b_i + 1$].
        \item [Z*6.2] Set [$b_{i+1} \gets a_i$].
    \end{description}
    \item [Z*7] For [$i$] From [$c$] To [$n$], execute the following steps:
    \begin{description}
        \item [Z*7.1] Set [$a_i \gets k_{max} - 1$].
        \item [Z*7.2] Set [$b_i \gets k_{max} - 1$].~~~$\blacksquare$
    \end{description}
\end{description}

It is trivial to see that Algorithm Z* does the same number of steps as Algorithm Z and is bound by the same number of operations. Algorithms V, W, X and Y can easily be reversed using the same logic too but, as previously argued, this is not deemed necessary as Algorithm Z* supersedes them.

\section{Problem E}
\label{sec:problem-e}

Problem E is the last problem that we define in this document and the most general one; any of the Problems A -- D can be reduced to Problem E and any algorithm that solves Problem E can also solve any other problem by simple adjustment of the parameter $K$. It consists in enumerating the partitions of a set, whose number of blocks are present in the set $K=\{k_1,k_2,k_3,\dots\}$. If the values of $k_i$ themselves are continuous, for example $\{k_{min},k_{min}+1,k_{min}+2,\dots,k_{max}\}$, then the problem is equivalent to Problem D for parameters $k_{min}$ and $k_{max}$.

Using the methods mentioned in Sections~\ref{sec:problem-b} and~\ref{sec:problem-c}, it can be shown that Problem E can easily be solved using Algorithm Z in constant amortized time by simple filtering of the enumerated partitions. This can be achieved using an additional boolean array $c$, where $c_i$ shows whether $i$ exists in the $K$ set, and can be created in time independent of $n$. Once again, however, in this section we present an algorithm for solving Problem E without backtracking, using a single (two-way) pass over the $a$ array; we call this Algorithm U.

Initially, regarding the parameter $K$, we sort it's array representation in ascending order so that $K=[k_{min},\dots,k_{max}]$. Then, we create the additional array $m$, where $m_i$ is equal to the lowest value in $K$ that is greater or equal to $i$. As a result, $m$ needs to hold at least $k_{max}$ elements. For example, if $K=[2,5]$, then $m=[2,2,5,5,5]$. This type of initialization is performed in linearithmic time in respect to $k_{max}$, which is independent of $n$ and, hence, the impact of the initialization in the amortized cost of each \texttt{next} call is zero. If $k_{max}$ is known in advance, it can also be done in linear time.

The difference of Problem E with Problem D can be shown with an example. Consider $K=[2,5]$ and the partition 011111. The next partition cannot be 011112 because it is a 3-partition and 3 is not in the $K$ set. Incrementing the second digit is also not possible because that would lead to 01112{-} and there is no suitable value for the LSD that leads to a valid partition; the minimum is 0 which leads to a 3-partition and the maximum is 3 which leads to a 4-partition. It can be shown that incrementing the third digit is possible in this case as it leads to 0112{-}{-}, which can hold a valid partition, e.g., the 5-partition 011234.

By considering the previous example we can induce a rule that dictates the first digit $a_c$ that can be incremented. After the index $c$ takes the value $a_c+1$, it is clear that the next partition cannot have less number of blocks than $t=\text{max}(a_1,a_2,\dots,a_c)+1$ and, as a result, the algorithm, using the $m_t$ value, determines the least amount of blocks that can be created; this process guarantees that the partition $a_{c+1} a_{c+2} \dots a_n$ will be either lexicographically minimal or invalid. The partition would be invalid if the remaining digits ($n-c$) are not enough to trigger an increase from $t$ number of blocks to $m_t$ number of blocks. In essence, the partition is (valid) lexicographically minimal if $n-c \ge m_t-t$. If the partition is invalid, the algorithm tries the next $c$ (the $c-1$ digit). In the edge case where $m_t=t$, it is implied that the single increment on the index $c$ results in a valid number of blocks and as a result the rest of the digits to the right can be filled with zeroes ($m_t-t=0$). Algorithm U is formulated below.

\begin{description}
    \item [Algorithm U] Accepts the array $a$ representing the previous partition, the auxiliary state variable $b$ representing the prefix maxima of array $a$ and the sorted arrays $k$ and $m$. Changes the array $a$ in-place to point to the next partition and updates the array $b$ to reflect the changes. No change is performed on $k$ or $m$.
    \item [U1] Set [$c \gets n + 1$], [$k_{max} = \text{max}(k)$], [$k_{min} = \text{min}(k)$].
    \item [U2] Do the following While [$a_c = k_{max} - 1$ Or $a_c > b_c$ Or $m_{t+1}-t>n-c$]
    \begin{description}
        \item [U2.1] Set [$c \gets c - 1$].
        \item [U2.2] Set [$t \gets \text{max}(a_c + 1, b_c)$].
    \end{description}
    \item [U3] If [$c = 1$] Report the end of enumeration.
    \item [U4] Set [$a_c \gets a_c + 1$].
    \item [U5] Set [$b_{c+1} \gets \text{max}(a_c, b_c)$].
    \item [U6] Set [$z \gets b_{c+1} + n - c - m_{b_{c+1}+1}$].
    \item [U7] Execute [$z$] times and as long as $c \le n$:
    \begin{description}
        \item [U7.1] Set [$a_c \gets 0$].
        \item [U7.2] Set [$b_{c+1} \gets b_c$].
        \item [U7.3] Set [$c \gets c + 1$].
    \end{description}
    \item [U8] For [$i$] From [$c$] To [$n$], execute the following steps:
    \begin{description}
        \item [U8.1] Set [$a_i \gets b_i + 1$].
        \item [U8.2] Set [$b_{i+1} \gets a_i$].~~~$\blacksquare$
    \end{description}
\end{description}

The process is similar to the other one pass algorithms and can be divided into 3 steps: find the index $c$ to increment (U2), switch direction and place as many zeroes as possible (U7) and, finally, fill up the rest of the $a$ array to guarantee the $m_t$-partition (U8). The steps U7 and U8 are almost identical to steps Z7 and Z8 but instead of $k_{min}$ we consider $m_t$ as shown earlier.

The number of enumerations performed in Problem E are
\begin{equation*}
    \sum_{k \in K} \mathcal{S}(n,k),
\end{equation*}
while in each one the probability of $c$ not being the LSD is
\begin{equation*}
    \frac{\sum_{k \in K} \mathcal{S}(n-1,k)}{\sum_{k \in K} \mathcal{S}(n,k)}
\end{equation*}
which for the reasons explained in Equations~\ref{eq:asymptotic-2} and~\ref{eq:limit-striling-sum} is asymptotically equal to
\begin{equation*}
    \frac{\mathcal{S}(n-1,\text{max}(K))}{\mathcal{S}(n,\text{max}(K))} = \frac{1}{\text{max}(K)}
\end{equation*}
and equivalent to the respective probability of Algorithm Z. As a result, the number of operations are asymptotically proportional to
\begin{equation*}
    \frac{1}{\text{max}(k)-1} = \frac{1}{k_{max}-1}.
\end{equation*}
Furthermore, Algorithm U uses arrays of maximum size $n$ and, thus, the extra memory required is proportional to $n$. It is easy to show that the maximum number of steps of the \texttt{next} method of Algorithm U is also linear in respect to $n$.

Regarding the reverse lexicographic version of Algorithm U, it is easy to produce Algorithm U* with the extra parameter $r$ which is, in essence, the inverted $m$ array; the value $r_i$ shows the maximum $k$ which is less than or equal to $i$:

\begin{description}
    \item [Algorithm U*] Accepts the array $a$ representing the previous partition, the auxiliary state variable $b$ representing the prefix maxima of array $a$ and the sorted arrays $k$, $m$ and $r$. Changes the array $a$ in-place to point to the next partition and updates the array $b$ to reflect the changes. No change is performed on $k$, $m$ or $r$.
    \item [U*1] Set [$c \gets n + 1$], [$k_{max} = \text{max}(k)$], [$k_{min} = \text{min}(k)$].
    \item [U*2] Do the following While [$a_c = 0$ Or $m_{t+1}-t>n-c$]
    \begin{description}
        \item [U*2.1] Set [$c \gets c - 1$].
        \item [U*2.2] Set [$t \gets \text{max}(a_c - 1, b_c)$].
    \end{description}
    \item [U*3] If [$c = 1$] Report the end of enumeration.
    \item [U*4] Set [$a_c \gets a_c - 1$].
    \item [U*5] Set [$b_{c+1} \gets \text{max}(a_c, b_c)$].
    \item [U*6] Set [$z \gets r_{b_{c+1} + n - c}$].
    \item [U*7] Set [$c=c+1$].
    \item [U*8] While [$b_i < z - 1$]
    \begin{description}
        \item [U*8.1] Set [$a_c \gets b_c + 1$].
        \item [U*8.1] Set [$b_{c+1} \gets a_c$].
    \end{description}
    \item [U*9] For [$i$] From [$c$] To [$n$], execute the following steps:
    \begin{description}
        \item [U*9.1] Set [$a_i \gets z-1$].
        \item [U*9.2] Set [$b_{i+1} \gets z-1$].~~~$\blacksquare$
    \end{description}
\end{description}

In conclusion, Algorithm U is the most general algorithm for solving the set partition enumeration problem. Its parameter $K$ dictates the desired number of blocks of the partitions and it can contain arbitrary values in $[1,n]$. It operates in constant amortized time per \texttt{next} call using, in the worst case, a single two-way pass over the restricted growth string array and is optimal under this condition.

\section{Summary}

This document brings attention to fundamental lexicographic set partition enumeration problems in a perspective that is suitable for learners and programmers. We approach the presentation of the problems in a progressive way that is easy to understand, and thoroughly explain the algorithms used to solve them. We show how these algorithms work and demonstrate their amortized constant time complexity. These algorithms are straightforward to implement in modern, high-level programming languages and don't have any complicated \textit{jump} logic or recursive structures. Table~\ref{tab:algorithms} presents a summary of the algorithms discussed in this document. Reference implementations of these algorithms in C++ are available online\footnote{https://github.com/gstamatelat/partitions-enumeration}.

\begin{table}
    \centering
    \setstretch{0.5}
    \begin{tabular}{cccccc}
        \hline\\
        Algorithm & Problem & Time & Space & Order & Iterations \\\\
        \hline\\
        V & A & $\displaystyle\sim 1$ & $\sim n$ & Lexicographic & $\displaystyle\mathcal{B}_n$ \\\\
        W & B & $\displaystyle\sim \left( 1 + \frac{1}{k-1} \right)$ & $\sim n$ & Lexicographic & $\displaystyle\mathcal{S}^+(n,k)$ \\\\
        X & C & $\displaystyle\sim \left( 1 + \frac{1}{k-1} \right)$ & $\sim n$ & Lexicographic & $\displaystyle\mathcal{S}(n,k)$ \\\\
        Y & C & $\displaystyle\sim \left( 1 + \frac{1}{k-1} \right)$ & $\sim n$ & Lexicographic & $\displaystyle\mathcal{S}(n,k)$ \\\\
        Z & D & $\displaystyle\sim \left( 1 + \frac{1}{k_{max}-1} \right)$ & $\sim n$ & Lexicographic & $\displaystyle\sum_{k=k_{min}}^{k_{max}} \mathcal{S}(n,k)$ \\\\
        Z* & D & $\displaystyle\sim \left( 1 + \frac{1}{k_{max}-1} \right)$ & $\sim n$ & Reverse Lex. & $\displaystyle\sum_{k=k_{min}}^{k_{max}} \mathcal{S}(n,k)$ \\\\
        U & E & $\displaystyle\sim \left( 1 + \frac{1}{\text{max}(K)-1} \right)$ & $\sim n$ & Lexicographic & $\displaystyle\sum_{k \in K} \mathcal{S}(n,k)$ \\\\
        U* & E & $\displaystyle\sim \left( 1 + \frac{1}{\text{max}(K)-1} \right)$ & $\sim n$ & Reverse Lex. & $\displaystyle\sum_{k \in K} \mathcal{S}(n,k)$ \\\\
        \hline
    \end{tabular}
    \caption{Summary of the algorithms discussed in this work.}
    \label{tab:algorithms}
\end{table}

This work can be extended by including various algorithms for the generation of random set partitions; in the general case a random partition with the required number of blocks inside an integer set $K$. Another line of future work is the restatement of Problems A, B, C, D and E for gray code enumerations instead of lexicographic.

\bibliographystyle{apalike}
\bibliography{main}

\end{document}